\documentclass[aps,twocolumn]{revtex4}

\usepackage{graphicx}
\input epsf

\begin{document}
\bibliographystyle{revtex}

\title{A Hybrid Model of Neutrino Masses and Oscillations:\\
Bulk Neutrinos in the Split-Fermion Scenario}

\author{Keith R. Dienes$^1$\footnote{E-mail address: {\tt dienes@physics.arizona.edu}},
$\,$Sabine Hossenfelder$^{1,2}$\footnote{E-mail address: {\tt sabine@physics.ucsb.edu}}}
\affiliation{$^1$Department of Physics, University of Arizona, Tucson, AZ 85721 USA\\
$^2$Department of Physics, University of California, Santa Barbara, CA  93106  USA}

\date{July 10, 2006}

\begin{abstract}
Higher-dimensional models of neutrino physics  
with one or more right-handed neutrinos in the bulk
have attracted considerable attention in recent years.
However, a critical issue for such models is to find a way
of introducing the required flavor dependence needed for
generating neutrino oscillations.
In this paper, we point out that a natural ``minimal''
framework that accomplishes this can be constructed by combining
the bulk-neutrino hypothesis for right-handed neutrinos
with the split-fermion scenario for left-handed neutrinos.
This combination leads to a unique flavor signature for
neutrino phenomenology which easily incorporates large flavor mixing
angles.
This hybrid scenario also has a number of additional important features.
For example, one previous difficulty of the split-fermion scenario applied
to neutrinos has been that the mass matrix is exponentially sensitive
to neutrino displacements within the brane.
However, in our hybrid scenario, the interactions between the brane and bulk
naturally convert this dependence from exponential to linear.
Another important feature is that our hybrid scenario provides its 
own natural regulator for Kaluza-Klein sums.
Thus, in our scenario, all Kaluza-Klein summations are manifestly finite,
even in cases with multiple extra dimensions.
But most importantly, our mechanism completely decouples the effective
neutrino flavor mixing angles from the sizes of the overlaps
between the neutrino wavefunctions within the brane.
Thus, we are able to obtain large neutrino mixing angles
even when these neutrinos have significant 
spatial separations and their overlaps vanish.
\end{abstract}

\maketitle

\newcommand{\newc}{\newcommand}

\newc{\gsim}{\lower.7ex\hbox{$\;\stackrel{\textstyle>}{\sim}\;$}}
\newc{\lsim}{\lower.7ex\hbox{$\;\stackrel{\textstyle<}{\sim}\;$}}
\def\half{{\textstyle{1\over 2}}}

\def\inbar{\,\vrule height1.5ex width.4pt depth0pt}

\def\IC{\relax\hbox{$\inbar\kern-.3em{\rm C}$}}
\def\IQ{\relax\hbox{$\inbar\kern-.3em{\rm Q}$}}
\def\IR{\relax{\rm I\kern-.18em R}}
 \font\cmss=cmss10 \font\cmsss=cmss10 at 7pt
\def\IZ{\relax\ifmmode\mathchoice
 {\hbox{\cmss Z\kern-.4em Z}}{\hbox{\cmss Z\kern-.4em Z}}
 {\lower.9pt\hbox{\cmsss Z\kern-.4em Z}}
 {\lower1.2pt\hbox{\cmsss Z\kern-.4em Z}}\else{\cmss Z\kern-.4em Z}\fi}

\def\beq{\begin{equation}}
\def\eeq{\end{equation}}
\def\beqn{\begin{eqnarray}}
\def\eeqn{\end{eqnarray}}


\section{Introduction and Overview}
 
Many physicists consider neutrino oscillations to be the first clear signature for 
physics beyond the Standard Model. With the most recent data from SNO and Super-Kamiokande,
there remains virtually no doubt about the existence of neutrino flavor oscillations. 
Indeed, the parameter space for neutrino mass differences
and mixing  angles has already become significantly constrained.

There remains, however, the paramount issue concerning how these experimentally determined
parameters can be accommodated within a theoretically motivated model.
While there have been numerous avenues that have been explored within recent years,
models with large extra dimensions    
have received considerable attention
ever since it 
was realized~\cite{add,ddg,wittlykk,antoniadis,rs}
that extra spacetime dimensions have
the potential to 
provide new, intrinsically geometric perspectives 
on the hierarchies
between various energy scales present within and beyond the Standard Model (SM).
These include the Planck scale~\cite{add,rs}, 
the GUT scale~\cite{ddg}, 
and the string scale~\cite{wittlykk,ddg},
as well as a possible scale for SUSY-breaking~\cite{antoniadis}.
Neutrino phenomenology has also been investigated within this higher-dimensional
context,
with the fundamental idea~\cite{addneut,ddgneut} being that since the 
right-handed neutrino is a Standard-Model singlet, it need not
necessarily be bound to the brane to which the other Standard-Model
particles are restricted.  The right-handed neutrino can therefore   
propagate into the bulk of extra spacetime dimensions and accrue
an infinite tower of Kaluza-Klein (KK) excitations, all of which will
then take part in neutrino oscillations~\cite{ddgneut}.  This idea, 
along with numerous variations, has spawned a relatively large 
literature investigating various aspects
of higher-dimensional neutrino phenomenology 
(see, {\it e.g.}\/, Refs.~\cite{addneut,ddgneut,many,ds}).

The challenge for these bulk-neutrino models, however, is to accommodate 
a suitable flavor structure
which can trigger the observed neutrino oscillations.
To date, two different directions have been explored in the literature.
 
One direction involves generalizing
the original idea of Refs.~\cite{addneut,ddgneut} by introducing
a separate right-handed bulk neutrino for each flavor of neutrino on the brane.
This process thereby extends flavor into the
bulk.  However, in such cases,
the brane/bulk couplings become arbitrary $3\times 3$ mixing matrices
whose parameters are in principle undetermined.
Moreover, the three bulk neutrinos
can in principle correspond to different
extra spacetime dimensions with
different radii.  Thus, one obtains a scenario
with many undetermined parameters governing neutrino masses
and mixing angles.  Indeed, to a large extent, the role of
the extra dimensions in such scenarios is reduced to providing
a set of Kaluza-Klein states which function 
(from a four-dimensional
perspective) as additional sterile neutrinos.  In contrast,
flavor neutrino
oscillations continue to be triggered through mixing angles
which are ultimately introduced by hand, just as in four dimensions.  
This therefore fails to
yield any further insight
regarding the structure or origins of the neutrino mixings.

A second, more ``minimal'' idea first advanced in Ref.~\cite{ds}
is to introduce only one bulk neutrino, and to provide this single bulk
neutrino with
flavor-universal couplings to all three brane neutrinos.
Thus, in so doing, one is essentially considering flavor
to be a feature internal to the Standard Model, a feature
which is restricted
to the brane and which therefore does not extend
into the bulk.
It is even possible to take the brane
theory to be flavor-diagonal, thereby avoiding
mixing angles completely; 
indeed, in the model of Ref.~\cite{ds},
the only flavor-sensitive feature that distinguishes these brane neutrinos
is their differing bare Majorana masses $m_i$.  
Nevertheless, one finds~\cite{ds} that sizable neutrino flavor oscillations
between the different neutrinos on the brane
arise as a result of their indirect mixings with the Kaluza-Klein modes
of the bulk right-handed neutrino.
Thus, in such a model, the flavor oscillations are entirely
``bulk-mediated'' and no mixing angles are needed at all:   
it is the presence of the higher-dimensional bulk which is completely
responsible for inducing the flavor oscillations on the brane.

Although it is possible to consider cases in which the couplings
between the brane neutrinos and the bulk neutrino are large,
it is convenient (and perhaps also phenomenologically necessary)
to consider a so-called ``perturbative'' limit in which these
couplings are relatively small.  In such cases, it is possible
to integrate out the effects of the Kaluza-Klein bulk neutrinos in 
order to obtain an effective
mixing matrix involving only the light modes in the theory.
Typically, these light modes consist of the three brane neutrinos as well as the
zero mode(s) of the bulk neutrino(s).
Because this effective mixing matrix completely encapsulates the resulting
flavor-dependence of the model, it inevitably lies at the
center of any comparison between a given theoretical neutrino model
and experimental data.  Indeed, experimentalists will
eventually provide a unique numerical mixing matrix which incorporates 
the entire observed phenomenology of neutrino oscillations.
It will then be the job of the theorist to solve the
``inverse'' problem of deducing the set of viable underlying neutrino
models which can lead to this matrix.

The non-minimal models with many bulk neutrinos
contain many mixing angles and can therefore give rise to relatively diverse
effective mixing matrices. 
However, one difficulty with the minimal model of Ref.~\cite{ds}
is that it leads to a rather rigid form  
for this effective mixing matrix.
Since the minimal model involves only one bulk neutrino,
it yields an effective $4\times 4$ mixing matrix 
which, as we shall see, has a texture of the form
\beq
       {\cal M} ~=~ \pmatrix{
      m_1 + X &  X  &  X  &  m  \cr
         X  &   m_2+X  &  X  &  m  \cr
         X  &   X  &  m_3 + X &  m \cr
       m  &  m &  m &  0 \cr}
\label{DSmassmatrix}
\eeq
where the first three rows/columns correspond
to the brane neutrinos and the final row/column corresponds
to the bulk zero mode.
Here $X$ and $m$ are parameters
associated with the bulk of the higher-dimensional theory and its
coupling to the brane,
while $m_i$ are the bare Majorana masses of the neutrinos on the brane.  
While we see that the bulk physics (through the term $X$) is responsible 
for yielding a non-diagonal $3\times 3$ brane mixing submatrix,
this matrix is relatively
rigid, with all off-diagonal entries forced to be exactly equal.
Indeed, a mixing texture of this form may have difficulty accommodating
observed neutrino oscillations of the correct sizes and magnitudes.

It is the purpose of this paper to propose an alternative
``minimal'' model which leads to a richer effective mixing matrix
without introducing additional parameters or sacrificing any
of the minimality of the single bulk-neutrino scheme.
Rather than incorporate a flavor structure through
differing bare Majorana masses $m_i$
on the brane as in Ref.~\cite{ds},
we shall instead incorporate a non-trivial flavor structure
within the context of the so-called ``split-fermion'' 
scenario~\cite{split}.
As is well known, the split-fermion scenario
represents an intrinsically higher-dimensional method
of explaining flavor hierarchies for 
the charged matter content of the Standard Model.
However, as we shall discuss below, this method faces certain
unique difficulties when attempting to address the flavor
structure of the neutrino sector. 
Thus, by combining the bulk-neutrino scenario with the split-fermion
scenario, we are able to obtain a 
``hybrid'' minimal model 
which has both strong theoretical motivations as well
as rich prospects for neutrino phenomenology.

\section{Split Fermions}

In the split-fermion  scenario~\cite{split},
the fermions of the Standard Model are located within a ``fat'' brane but centered
around different positions within the brane.  Their spatial extent within the fat brane
is modeled by a Gaussian distribution with a typical width $\sigma$, where $\sigma$ is
approximately one order of magnitude smaller
than the width of the brane. 
The gauge fields and the Higgs are assumed to be equally distributed
over the width of the brane. 
To obtain the effective four-dimensional couplings between 
different particles, the extra dimensions have to be
integrated out. 
Since the overlap of the Gaussian wavefunctions of the particles can
be very small, the resulting effective couplings can be extremely 
suppressed.  Indeed, it is easy to obtain a suppression 
by thirty orders of magnitude by delocalizing particles at locations
which are separated by about 
ten times their Gaussian widths. 
 
This setup yields  a solution to several problems. 
It can be used to suppress the proton decay, 
to suppress flavor-changing operators, or to
explain the observed mass hierarchy in the lepton sector. 
Many possible configurations of particles inside the brane 
are possible;  some of these are discussed in
Ref.~\cite{othersplits,Harnik,maps}.

Relative shifts between left-handed Standard-Model doublets and right-handed
singlets have been successfully used to address the mass hierarchy of quarks and charged leptons. 
But for the same reason the split-fermion  scenario is easy to implement for charged 
fermions, it leads to difficulties when one tries to incorporate neutrinos~\cite{Splitnus}. 
Experiments suggest with an increasing degree of certainty
that neutrino mixing is maximal in the case of $\nu_\tau \leftrightarrow \nu_\mu$ and
almost maximal in the case of $\nu_e \leftrightarrow \nu_\tau$.
The common textures of the zeroth-order 
neutrino mass matrix that are compatible with such almost-bimaximal mixing can be
classified by the hierarchy type into different categories --- normal,
inverted, or degenerate.
However, all of these situations 
require several  entries to be of the same order of magnitude.  

Unfortunately, in the split-fermion  scenario, the coupling between any
two particles is extremely sensitive
to their relative distance within the brane.  The observed neutrino
mixing can therefore be achieved only by carefully 
choosing the central locations of the neutrinos inside the brane. In addition, 
suitable positions for the 
charged leptons must be found in order to avoid
excessive flavor-changing processes.  Together, 
these requirements put severe constraints on the allowed fermion
locations~\cite{Splitnus}.
While all experimental constraints can ultimately be accommodated,
the resulting particle map is extremely sensitive to small perturbations.  
Thus, even though these models
reduce the amount of fine-tuning for the Yukawa couplings, a
considerable
degree of fine-tuning continues to be necessary.

There are, of course, various options for ameliorating this situation.
For example, 
introducing Majorana neutrinos rather than Dirac neutrinos
has been proposed~\cite{othermasssplit} 
as a method of
achieving reasonable neutrino masses within the split-fermion  scenario. 
The neutrino sector of the split-fermion scenario has also recently been 
examined~\cite{Moreau}
within the context of the Randall-Sundrum model~\cite{rs}.
 
In this paper, we will consider a different approach towards the neutrino 
sector by asking what consequences arise if the two scenarios --- 
split fermions and bulk neutrinos --- are examined in a joined framework. 
In this way, we shall be combining the strengths of each individual
scenario:  the split-fermion mechanism will trigger the flavor
structure of the neutrino sector, while the bulk sector will
ameliorate the fine-tuning issues and naturally lead to large
mixing angles. 
Indeed, it was already noted in Ref.~\cite{ddgneut} that the combination of
having brane neutrinos at different locations in the extra dimension
would trigger a higher-dimensional
seesaw mechanism, although this idea was not pursued and
no consistent framework was provided. 
In this paper, we shall see that split fermions provide a natural
context for this phenomenon, leading to
flavor-dependent 
mixing properties which can be used to distinguish between the brane neutrinos.

There are also other benefits to combining these scenarios.
For example, as we have mentioned,
one difficulty of the split-fermion scenario applied
to neutrinos has been that the mass matrix is exponentially sensitive
to neutrino displacement within the brane.
However, in our hybrid scenario, we shall see that the interactions between the brane and bulk
naturally convert this dependence from exponential to linear.
Thus, the previous exponential sensitivity is entirely eliminated. 
Another important feature is that our hybrid scenario provides its
own natural regulator for Kaluza-Klein sums.
Thus, in our scenario, all such Kaluza-Klein summations are manifestly finite,
even in cases with multiple extra dimensions.
As we shall see, this arises because 
the heavy Kaluza-Klein modes with wavelengths that are small compared with the
widths of the brane Gaussians
will average out when folded with the Gaussian distribution. This
thereby
eliminates the necessity of introducing a cutoff by hand.

But most importantly, we shall find that
our mechanism completely decouples the effective
neutrino flavor mixing angles from the magnitudes of the 
neutrino overlaps within the brane.  
Thus, we are able to obtain large neutrino mixing angles
between neutrinos, even when these neutrinos have rather large spatial separations
within the brane.

\section{The hybrid framework}
\label{model}

We now describe the framework for our hybrid model.
Clearly, parts of this model will be similar to the model
considered in Ref.~\cite{ds}.  For concreteness, 
we will work in this section  with one bulk neutrino and one 
compactified extra dimension of radius $R$,
although
the model can be  trivially extended to more bulk fields and to more
additional dimensions.

The primary features of this model can be summarized as follows.
We imagine a single extra dimension of radius $R$, as well as a brane
of width $R'\ll R$.  
\begin{itemize}
\item On the brane, we introduce
       $n_{\rm f}$ left-handed neutrinos $\nu_\alpha$;
       these are our flavor eigenstates.
       We assume that these left-handed neutrinos have
       no corresponding Majorana masses and are restricted 
       to lie within the brane.
\item In the bulk, we introduce a single Dirac fermion $\Psi$ which
     does not carry any flavor indices and is therefore
    completely flavor-neutral. 
\item  For the sake of minimality, we assume that the 
         left-handed neutrinos do not mix or couple to each other directly.
      Thus, all mixing angles on the brane are set to zero.  
\item We introduce a single flavor-blind brane/bulk 
      coupling $M_*$ between each of the brane neutrinos $\nu_\alpha$
      and the bulk neutrino $\Psi$. 
\item Finally, each of the active brane neutrinos $\nu_\alpha$
      is assumed to be centered around a different transverse location
      within the brane.  It is this feature, and only this feature,
      which is introduces a flavor-dependence 
      into our model.
\end{itemize}

We shall take our spacetime coordinates as $x^A\equiv (x^\mu,y)$
where $x^\mu$ are the four uncompactified coordinates 
on the brane
and $y$ is the coordinate of the fifth dimension perpendicular to the brane.
Given the above assumptions, the action for our model
then takes the form
$ {\cal S} = {\cal S}_{\nu} + {\cal S}_{\rm b} + {\cal S}_{\rm c} + {\rm h.c.} $
where $S_{\nu}$ describes the physics on the brane,
$S_{\rm b}$ describes the physics in the bulk,
and $S_{\rm c}$ describes the brane/bulk couplings.
These individual contributions take the form  
\beqn
  {\cal S}_{\nu} &=& 
       \int d^4 x \,dy~ \sum_{\alpha=1}^{n_{\rm f}} 
        \nu^\dag_\alpha (x,y) {\rm i} {\bar\sigma}^\mu \partial_\mu \nu_\alpha (x,y)
         \nonumber\\
  {\cal S}_{\rm b} &=& \int d^{4} x \,dy  \,
     \bar \Psi(x,y)    \, {\rm i}{\Gamma}^A \partial_A\,  \Psi(x,y)   \nonumber\\
  {\cal S}_{\rm c} &=&  
  \int d^4 x \,dy~
     \sum_{\alpha=1}^{n_{\rm f}}  
        M_* \, \nu^\dag_\alpha (x,y)  \left[\psi_+(x,y)  + 
      \bar \psi_-(x,y) \right]\nonumber\\  
\label{klag}
\eeqn
where $\Gamma$ represent the five-dimensional Dirac matrices.
Here the index $\alpha=1,..,n_{\rm f}$ runs over all $n_{\rm f}$ flavors
on the brane.

In the above, we have chosen to  
work in the Weyl basis in which the Dirac fermion $\Psi$ can be decomposed
into two two-component spinors:  $\Psi = (\psi_+,\bar\psi_-)^T$. 
The bar on $\bar\psi_-$ indicates that $\bar\psi_-$ transforms as a
different Lorentz representation than $\psi_+$. 
We refer to $\psi_+$ and $\bar\psi_-$ 
as left-handed and right-handed components, respectively.
We have also written
$\sigma^\mu=(1,-\sigma^i), \bar\sigma^\mu=(1,+\sigma^i)$ where $\sigma^i$ are
the Pauli matrices.

Given the action in Eq.~(\ref{klag}), 
the next step is to compactify the fifth dimension in order to
obtain an effective four-dimensional theory.
To do this, we shall make the following assumptions.
First, we shall identify $y\approx y+2\pi R$ and  
impose the orbifold relations
$\psi_{+}(-y)=\psi_{+}(y)$ and $\bar\psi_{-}(-y)=-\bar\psi_{-}(y)$,
where $y$ is the coordinate of the fifth dimension.  
Thus, $\psi_+$ and $\bar\psi_-$ have the Kaluza-Klein
mode expansions 
\beqn
     \psi_+(x,y) &=&  {1\over\sqrt{\pi R}} \, \psi_+^{(0)}(x) \nonumber \\
     && ~~~~~ +  {1\over \sqrt{2 \pi R}} \, \sum_{n > 0}   
           \psi_+^{(n)}(x)\,\cos \left( {ny\over R}\right) \nonumber\\
     \bar\psi_-(x,y) &=& {1\over \sqrt{2 \pi R}}  \, \sum_{n> 0}
         \bar\psi_-^{(n)}(x)\,\sin \left( {ny\over R}\right)~.
\label{KKdecomp} 
\eeqn  
Note that as a result of our orbifold boundary conditions,
$\bar \psi_-$ does not have a zero mode.
We shall also take our brane neutrinos to have wavefunctions
of the form
\beqn
   \nu_\alpha (x,y) ~=~ G(y-y_\alpha,\sigma) \, \nu_\alpha(x) ~ 
\label{nudecomp}
\eeqn
where $\nu_\alpha(x)$ denotes the usual four-dimensional part of the
fermion wavefunction on our brane and
where $G(y,\sigma)$ denotes a normalized 
Gaussian wavefunction centered around location $y=0$  with width $\sigma$:
\beqn
   G(y,\sigma) = \frac{1}{\sqrt{\sigma}} 
   \exp \left(- \frac{\pi}{2} \frac{y^2}{\sigma^2} \right)~.
\eeqn
Note that for simplicity, we are assuming that each brane
neutrino has the same width $\sigma \ll R$.

Inserting the wavefunctions in Eqs.~(\ref{KKdecomp}) and (\ref{nudecomp}) into the
action (\ref{klag}) and 
integrating over the fifth dimension, we obtain the effective
four-dimensional actions:
\beqn
  {\cal S}_{\nu} &=&  
\int d^4 x ~ \sum_{\alpha=1}^{n_{\rm f}}   
     \nu^\dag_\alpha  {\rm i} {\bar\sigma}^\mu \partial_\mu \nu_\alpha  \nonumber \\
   {\cal S}_{\rm b} &=& \int d^4 x ~\biggl\lbrace
     {\bar\psi_-}^{\dag(0)} {\rm i}{\sigma}^\mu\partial_\mu \bar\psi_-^{(0)}  \nonumber \\
     && ~~+  \sum_{n > 0} \left[
     \psi_+^{\dag(n)} {\rm i} {\bar\sigma}^\mu\partial_\mu {\psi_+}^{(n)}
     +{\bar\psi_-}^{\dag(n)} {\rm i}{\sigma}^\mu\partial_\mu \bar\psi_-^{(n)} \right]  \nonumber\\
     &&
    ~~+ \sum_{n > 0}\, \left({n\over R}\right)
          \left\lbrack
          \psi_+^{\dag (n)}  \bar\psi_-^{(n)}
         +   \bar\psi_-^{\dag (n)}  \psi_+^{(n)} \right\rbrack \biggr\rbrace  
	  \nonumber \\
  {\cal S}_{\rm c}  &=&  
       \int d^4 x~  \sum_{\alpha=1 }^{n_{\rm f}}  
     \nu^\dag_\alpha \biggl\lbrace m \bar\psi_-^{(0)} \nonumber \\
 &&  ~~+   \sum_{n > 0} \left( m^\alpha_{n,+} \,\psi_+^{(n)} + 
          m^\alpha_{n,-} \, \bar\psi_-^{(n)} \right)
                   \biggr\rbrace  ~. 
\label{tglag3}
\eeqn
Note that the fields are now functions of $x^\mu$ 
only, and $m$ and $m^\alpha_{n,\pm}$
are volume-suppressed $n$-dependent brane/bulk
couplings resulting from the rescaling of the 
individual $\psi_+^{(n)}$, $\bar\psi_-^{(n)}$ Kaluza-Klein modes:  
\beqn
    m &\equiv&  M_* \sqrt{ \sigma\over 2\pi R}   \nonumber\\
    m_{n,+}^{\alpha} &\equiv&   \sqrt{2} m\, 
      \exp \left( - \frac{n^2}{2 \pi} \frac{\sigma^2}{R^2} \right)
    \cos \left( \frac{n}{R} y_\alpha \right) \nonumber\\
     m_{n,-}^{\alpha} &\equiv & \sqrt{2}  m\,
     \exp \left( - \frac{n^2}{2 \pi} \frac{\sigma^2}{R^2} \right)
    \sin \left(  \frac{n}{R} y_\alpha \right)   ~.
\label{couplings3}
\eeqn
As we see, the brane/bulk coupling 
is suppressed by a volume factor $\sqrt{\sigma/R}$.
Moreover, for non-shifted brane neutrinos with $y_\alpha=0$,
we see that the couplings $m^\alpha_{n,-}$ actually vanish.
Thus, it is only the possible displacement of the antisymmetric
fermion modes $\bar\psi_-^{(n)}$ away from $y_\alpha=0$ which
permits them to couple to the bulk zero mode.

Given the Lagrangian in Eq.~(\ref{tglag3}), we immediately see that
the SM flavor-eigenstate neutrinos
$\nu_\alpha$ on the brane will mix with the entire tower of Kaluza-Klein states
of the higher-dimensional $\Psi$ field, even though they do not mix
directly with each other.
Defining
\beq
        {\cal N}^T ~\equiv~ (\nu_\alpha,  {\psi}_+^{(0)},  \psi_+^{(1)}, 
                  {\bar \psi}_-^{(1)}, \psi_+^{(2)}, 
                  {\bar \psi}_-^{(2)},  ...)~,
\label{calNdef}
\eeq
we see that the 
mass term of the Lagrangian~(\ref{tglag3})
takes the form ${\cal N}^T {\cal M} {\cal N}$
where the mass matrix ${\cal M}$ is given by
\beq
      {\cal M} ~=~ 
        \pmatrix{
         0 &       m    &   \ldots & m^\alpha_{n,+}   &  m^\alpha_{n,-} &  \ldots \cr
             m^T    &  0   &   \ldots  &   0   &   0               &   \ldots \cr
       \vdots & \vdots & \ddots & \vdots &  \vdots & \ldots \cr   
 m_{n,+}^{\alpha T}   &  0   &   \ldots  &  0   &   n/R       & \ldots \cr
 m_{n,-}^{\alpha T}  &   0   &   \ldots  &  n/R  &   0  & \ldots \cr
       \vdots & \vdots & \vdots & \vdots &  \vdots & \ddots \cr }~.  
\label{oldmatrixp}
\eeq
Of course, in this notation, the first row/column represents $n_{\rm f}$ 
independent rows/columns, one for each flavor neutrino on the brane.
The fact that this mass matrix is non-diagonal implies that 
each of the $n_{\rm f}$ brane neutrinos will 
undergo oscillations with the bulk KK modes.  
Through these mixings with the bulk KK neutrinos,
the brane neutrinos will thereby 
undergo effective flavor oscillations with each other.

The next step is to determine the eigenvalues and eigenvectors of the 
mass matrix~(\ref{oldmatrixp}).
We can begin this task as follows.
Defining the eigenvector 
$v=(v^\alpha,v^{(0)},v^{(1)}_+,v^{(1)}_-,...,v^{(n)}_+,v^{(n)}_-,...)$,
we see that the eigenvalue equation ${\cal M}v = \lambda v$ 
yields the simultaneous equations:
\beqn
  m v^{(0)} + \sum_{n>0} \left( m_{n,+}^\alpha v_+^{(n)} +  m^\alpha_{n,-}  v_-^{(n)} \right) &=& \lambda v^\alpha \nonumber\\
  m \sum_\alpha v^\alpha &=& \lambda v^{(0)} \nonumber\\
  \sum_\alpha m^\alpha_{n,+} v^\alpha + \frac{n}{R} v_-^{(n)} &=& \lambda v_+^{(n)} \nonumber \\
  \sum_\alpha m^\alpha_{n,-} v^\alpha + \frac{n}{R} v_+^{(n)} &=& \lambda v_-^{(n)} ~.
\label{simulteqs}
\eeqn
Combining the last two of these equations yields 
a partial solution for $v_\pm^{(n)}$,
\beq
v_\pm^{(n)}= 
\frac{1}{\lambda^2 - (n/R)^2} \sum_\alpha v^\alpha 
           \left( m^\alpha_{n,\pm } \lambda + m^\alpha_{n,\mp} {n\over R}\right)~,
\label{interstep}
\eeq
and inserting this into 
the top line of Eq.~(\ref{simulteqs})
along with the definitions in Eq.~(\ref{couplings3}) yields
the relation
\beqn
  && \hskip -0.30 truein \lambda  v^\alpha ~=~ m v^{(0)} + 2 m^2 \sum_{n>0}  
            \frac{\exp \left( - \frac{n^2}{2 \pi} \frac{\sigma^2}{R^2} \right)}
                 {\lambda^2 - (n/R)^2} 
        \times \nonumber \\
  &&   \hskip -0.3 truein \sum_\beta v^\beta \left( \lambda \cos \left[ \frac{n}{R} (y_\alpha-y_\beta) \right] + 
  \frac{n}{R} \sin \left[ \frac{n}{R} (y_\alpha+y_\beta) \right] \right)~.  
\label{7ev} 
\eeqn
Thus, we see that the task of solving the 
remaining equations 
has now been reformulated as an equivalent 
$(n_{\rm f}+1) \times (n_{\rm f}+1)$-dimensional
eigenvalue problem 
of the form
\beq
     \widetilde{{\cal M}}(\lambda) \, \widetilde{v} ~=~ \lambda \, \widetilde{v} ~ 
\label{8ev}
\eeq
where $\widetilde{v} = (v^\alpha, v^{(0)})$
and where  
$\widetilde{{\cal M}}(\lambda)$
is a $\lambda$-dependent 
``reduced'' mass matrix of the form
\beq
     \widetilde{{\cal M}}(\lambda) ~=~ 
             \pmatrix{
              m_{\alpha \beta}(\lambda)    & m  \cr
                   m^{T}          & 0  \cr} ~.
\label{tildeM} 
\eeq
Here $m_{\alpha \beta}\equiv (M_\ast^2 \sigma/\pi) f_{\alpha\beta}$, 
where the dimensionless coefficients $f_{\alpha\beta}(\lambda)$ are given by
\beqn
   && \hskip -0.3 truein f_{\alpha \beta}(\lambda) ~=~ 
   \sum_{n>0}  \frac{\exp \left( - \frac{n^2}{2 \pi} \frac{\sigma^2}{R^2} \right)}{\lambda^2 R^2 - n^2} 
       \times \nonumber\\ 
       && 
    \left( \lambda R \cos \left[ \frac{n}{R} (y_\alpha-y_\beta) \right] + 
   n \sin \left[ \frac{n}{R} (y_\alpha+y_\beta) \right]  \right) ~.
\label{fdef}
\eeqn
Note that for numerical purposes, 
Eq.~(\ref{8ev}) provides a particularly 
useful simplification of the original infinite-dimensional
eigenvalue problem.
Solving Eq.~(\ref{8ev}) 
for $\lambda$ and $\widetilde{v}$ and 
inserting these results
into Eq.~(\ref{interstep}) then enables one to obtain the full 
infinite-dimensional eigenvector $v$. 
 
Thus far, we have not made any approximations.
However, in order to proceed analytically, we
shall now take the so-called ``perturbative'' limit $m R \ll 1$ in which
the brane/bulk coupling is small compared with the compactification scale.
Introducing this limit is one way of ensuring that 
the brane sector of our theory (or, as we shall see, a subset of that sector) 
will experience  little or no net loss of
probability into 
the ``sterile'' tower of bulk excitations. 
In this limit (see, {\it e.g.}\/, Ref.~\cite{ddgneut}), 
the eigenvalues of the excited KK neutrino modes will be approximately $\pm n/R$,
corresponding to the bulk/bulk sector of the mixing matrix, while
the remaining $n_{\rm f}+1$ eigenvalues 
will be much smaller than the masses of the excited modes. 
Specifically, we see from the form of the reduced mass matrix in Eq.~(\ref{tildeM}) that
the eigenvalues of these remaining light modes
will scale either as $\lambda \sim M_\ast^2 \sigma$ or as $\lambda \sim m$.
However, the assumption that $mR\ll 1$ implies that $\lambda R\ll 1$ in either case:
if $\lambda\sim m$ then clearly $\lambda R\sim mR\ll 1$, while if $\lambda\sim  M_\ast^2 \sigma$
then
\beq
                 \lambda R ~\sim ~ M_\ast^2 \sigma R ~\sim~ (m R)^2 ~\ll~1~.
\eeq
Thus in either case we can approximate
\beq
    f_{\alpha \beta}  ~\approx 
      ~ - \,\sum_{n>0} {1\over n} \, \exp \left( -\frac{n^2 \sigma^2}{\pi R^2} \right)
    \, \sin \left[ \frac{n}{R} (y_{\alpha} + y_{\beta}) \right]~
\label{interm}
\eeq
for these remaining light modes.
As we shall demonstrate in Appendix~A, this result may also be derived
through a direct perturbative diagonalization of our original mass matrix.

Note that the only case in which it is not justified to pass from 
Eq.~(\ref{fdef}) to Eq.~(\ref{interm}) occurs when $y_2$ is
exactly equal to $- y_1$, for in this case the second term 
in Eq.~(\ref{fdef}) vanishes exactly for all $n$ and is thus 
always smaller than the first.
However, since $\lambda$ is significantly suppressed for the light modes
we are considering,
we see that the resulting value of $f_{\alpha\beta}$ in this case is significantly 
smaller than it is for other values of $y_1$ and $y_2$.
The approximation in Eq.~(\ref{interm}), which sets 
this value of $f_{\alpha\beta}$ to zero, 
therefore continues to be roughly accurate even in this case, indicating 
a strongly suppressed
value for $f_{\alpha\beta}$ in the region near $y_2= -y_1$.

It is relatively straightforward to evaluate the KK sum in Eq.~(\ref{interm}),
especially since the presence of the Gaussian width for the brane neutrinos
serves as a natural regulator for the KK sum which renders it manifestly
finite.
Introducing the variable $k=(\sigma/R)n$ and noting that
$\sigma\ll R$,
we see that the sum can be approximated as an integral, yielding
\beqn
   f_{\alpha \beta}  &\approx&  -\, \int_0^\infty \frac{{\rm d} k}{k}\,  
      \exp (-k^2/\pi) \, \sin \lbrack k (y_\alpha+y_\beta)/\sigma \rbrack \nonumber\\
   &=&  - {\pi\over 2}\,   
   {\rm Erf}\left[ {\sqrt{\pi}\over 2\sigma} (y_\alpha+y_\beta)\right] ~. 
\label{effmassmatrix}            
\eeqn
Note that the error function is an odd function, 
${\rm Erf}\,(-x)= -{\rm Erf}\,(x)$,
with approximate behavior
\beq
           {\rm Erf}\,(x)~\approx ~ \cases{
                     x  &  for $|x|\lsim 1$ \cr
                   {\rm sign}(x) &  for $|x|\gsim 1$~.}
\eeq
Thus, defining the 
rescaled (dimensionless) 
displacements $\widetilde y_\alpha \equiv \sqrt{\pi} y_\alpha/(2\sigma)$,
we conclude that for the light modes in our theory, 
our reduced
$(n_{\rm f}+1)\times 
(n_{\rm f}+1)$-dimensional
mass matrix in Eq.~(\ref{tildeM}) takes the form 
\beq
         \widetilde M ~\approx~ \pmatrix{  m_{\alpha\beta} & m \cr 
                                       m^T & 0 \cr} 
\label{tildeM2}
\eeq
with  
\beq
      m_{\alpha\beta} ~\approx~  -{M_\ast^2 \sigma\over 2} \times
        \cases{
          \widetilde y_\alpha+\widetilde y_\beta & 
             for $|\widetilde y_\alpha+\widetilde y_\beta| \lsim 1$\cr
          {\rm sign}(\widetilde y_\alpha+\widetilde y_\beta) & 
             for $|\widetilde y_\alpha+\widetilde y_\beta| \gsim 1$~.}
\label{mresult}
\eeq

We see from Eq.~(\ref{mresult}) that 
when the neutrino displacements are of the same order as their Gaussian widths,
the entries of this mass matrix depend {\it linearly}\/ rather than exponentially
on the displacements!
This is an important observation, implying that
the usual exponential sensitivity to the 
fermion displacements has been eliminated
in our hybrid scenario, with the interaction with the bulk neutrino
serving to convert this sensitivity from exponential to linear. 
Moreover, we also observe the interesting fact that 
the interactions with the bulk neutrino actually remove {\it all}\/ 
sensitivity to the neutrino displacements when these displacements
become significantly larger than their Gaussian widths,
with mass-matrix values saturating
near the universal value $m_{\alpha \beta}\approx -\half M_\ast^2 \sigma$ 
in such cases.
Thus, in such cases, all flavor dependence is lost.

We also notice from Eq.~(\ref{mresult}) that our effective
mixings on the brane are not translationally invariant within
the brane, since they depend not on the difference $y_\alpha-y_\beta$,
but rather on the {\it sum}\/ $y_\alpha+y_\beta$.
Indeed, our off-diagonal mixing terms cancel only for $y_\alpha= -y_\beta$
rather than $y_\alpha=y_\beta$.
Moreover, since this dependence is 
on the sum rather than the difference,
the net distance from the origin at $y=0$ becomes an important quantity,
determining when the error function passes from its linear to constant regime. 
Once again, however,
these features represent the effects 
of the mixings with the bulk neutrino modes, for it is the KK orbifold
relations just above Eq.~(\ref{KKdecomp}) 
which implicitly select the origin $y=0$ as a special point, thereby breaking translational
invariance within the brane.

These features imply that it is possible to obtain  large effective flavor mixing
angles for the active neutrinos on the brane {\it even when 
the overlaps between these neutrinos vanish}\/.
In other words, large neutrino overlaps are no longer needed for 
large effective mixing angles.

To illustrate this behavior more concretely, 
let us consider a simplified two-flavor example in which we have only one
effective mixing angle.
Given the result in Eq.~(\ref{mresult}), we may identify this effective
mixing angle $\theta$ through the relation
\beq
       \tan 2\theta ~=~ {2\, {\rm Erf}\,(\widetilde y_1+\widetilde y_2) \over 
            {\rm Erf}\,(2 \widetilde y_2) - 
            {\rm Erf}\,(2 \widetilde y_1) }~.
\label{effangle}
\eeq
For a given fixed value of $\widetilde y_1$ (the position of the first neutrino), 
we can now examine how this effective mixing angle depends on $\widetilde y_2$ (the position
of the second neutrino).

There are several cases to consider.
If $|\widetilde y_1|\ll 1$ (so that the first neutrino is centered very close to the origin), 
we find that $\tan 2\theta\to 2$ as 
$\widetilde y_2\to \pm \infty$.  Thus, 
we find that $\sin^2 2\theta\to 4/5$ as
$\widetilde y_2\to \pm \infty$,  which represents 
nearly maximal mixing.  Note that this result applies
even though the second neutrino is being pulled relatively far from the first.
Indeed, one can verify the mixing remains nearly maximal in 
this case {\it regardless}\/ of the value of $\widetilde y_2$. 

By contrast, if $|\widetilde y_1|\gsim 1$, 
our asymptotic behavior changes.
For concreteness, let us take $\widetilde y_1$ positive  (otherwise
we can change all $y\to -y$ without altering the results).
In this case, Erf$\,(\widetilde y_1)\approx 1$ and we see from 
Eq.~(\ref{effangle}) that 
$\tan 2\theta\to \infty$ as $\widetilde y_2\to \infty$
while $\tan 2\theta\to 2$ as $\widetilde y_2\to -\infty$.
Thus, asymptotically, we have relatively large mixing angles in either case.
However, near $\widetilde y_2 \approx -\widetilde y_1$, we find that
our mixing angle passes through zero and changes sign.
Thus, there is a region near $\widetilde y_2 \approx -\widetilde y_1$
for which smaller mixing angles can be realized 
without exponential fine-tuning.

Finally, for $|\widetilde y_1|\lsim 1$, the resulting behavior interpolates
between the two behaviors described
above.  Again taking $\widetilde y_1>0$ for
simplicity, we find that $\sin^2 2\theta$ asymptotes to values
between $1/2$ and $4/5$ as $\widetilde y_2\to -\infty$,
and between
$4/5$ and $1$ as $\widetilde y_2\to +\infty$.
However, as in all cases, the effective mixing angle
hits zero at $\widetilde y_2=-\widetilde y_1$.

\begin{figure}[ht]
\centerline{
   \epsfxsize 3.7 truein \epsfbox {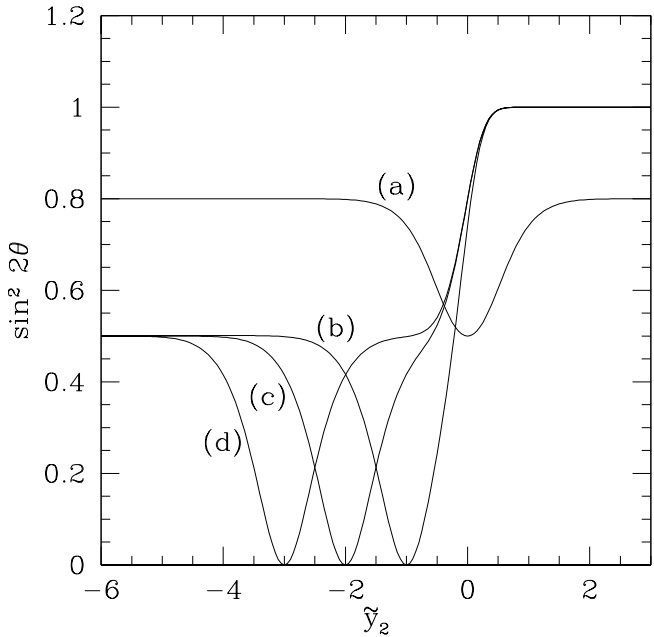}
    }
\caption{Effective flavor mixing angle $\sin^2 2\theta$ as a function of neutrino positions 
      $(\widetilde y_1,\widetilde y_2)$
       in the case of two active flavors on the brane.
     We plot $\sin^2 2\theta$ as a function of $\widetilde y_2$ 
     for $\widetilde y_1=0,1,2,3$ [curves (a) through (d) respectively].
    In each case, we see that large mixing angles are fairly common,
    with small mixing angles emerging only near $\widetilde y_2 \approx - \widetilde y_1$.  }
\label{figure}
\end{figure}
This behavior is shown in Fig.~\ref{figure} for $\widetilde y_1=0,1,2,3$.
In each case, we observe that large mixing angles dominate,
arising in most regions of the parameter space.  Nevertheless,
we see that small and intermediate mixing angles are 
easy to achieve 
without significant fine-tuning
in the region near $\widetilde y_2 = -\widetilde y_1$.

As indicated in the Introduction,
the flavor mixing in this model is significantly different
from that of an earlier model~\cite{ds} in which all brane neutrinos are
located at $\tilde y=0$ with zero width, distinguished
only through their different ``bare'' Majorana masses $m_i$ on the brane.
In such a model, the brane neutrinos couple to only the even bulk modes
$\psi_+^{(n)}$,
whereupon we see (following the same analysis as performed here)
that the effective $(n_{\rm f}+1)\times 
(n_{\rm f}+1)$ ``reduced'' mixing  
matrix $\widetilde {\cal M}(\lambda)$ takes the form shown in Eq.~(\ref{DSmassmatrix}),
where 
\beq
         X~=~  -{m^2\over \lambda} + \pi m^2 R \cot (\pi R\lambda)  ~. 
\eeq
The important point here is that $X$ is completely flavor-independent.
Thus all of the off-diagonal entries in the effective mass matrix 
are identical, leading to a relatively rigid flavor structure.

Thus far, we have focused on the case with only a single
extra spacetime dimension.
It is therefore worth understanding how the relevant mass scales in the 
mass matrix~(\ref{tildeM2}) are altered if we 
consider the analogous situation with $\delta>1$ 
extra dimensions.
With $\delta$ extra dimensions,
our three-brane has a ``width'' in $\delta$ directions,
so that our brane neutrinos are $\delta$-dimensional Gaussians.
Tracing our previous steps, this implies that the brane/bulk 
coupling $m$ defined in Eq.~(\ref{couplings3})
will accrue a further volume factor of $\sqrt{\sigma/R}$ for each
additional spacetime dimension, so that we can generally replace
\beq
             m ~~\to~~ m (\sigma/R)^{(\delta-1)/2} \sim M_\ast (\sigma/R)^{\delta/2}~.
\label{replace}
\eeq
On the other hand, 
it is shown in Appendix~B that the entries of $m_{\alpha \beta}$ continue to scale as
$M_*^2 \sigma$ regardless of the number or radii of extra dimensions.
Thus, for general $\delta$, we obtain
an effective $(n_{\rm f}+1)\times (n_{\rm f}+1)$-dimensional
mass matrix whose entries scale as
\beq
          \tilde M ~\sim~ M_\ast \, \pmatrix{
                      M_\ast \sigma\, f_{\alpha\beta} &  (\sigma/R)^{\delta/2}\cr
                      (\sigma/R)^{\delta/2} & 0 \cr}~
\label{tildeMdelta}
\eeq
where $f_{\alpha\beta}$ denotes a $\delta$-dependent
set of flavor-dependent coefficients of order one.
We shall not pursue a detailed calculation of these coefficients for $\delta>1$, 
and shall instead rely on our previous results 
for $\delta=1$
as a general illustration of the potential possibilities for the
flavor dependence.
 
These observations are  especially relevant for determining
the neutrino mass eigenvalues resulting from our mass matrices.
Indeed, one of the unique
features of these higher-dimensional scenarios involving 
right-handed neutrinos in the bulk
is that it is the {\it same}\/ mass matrix that determines not only
the neutrino flavor mixings  
but also the neutrino mass eigenvalues.  
This is different from the traditional four-dimensional framework
in which couplings to heavy right-handed  
neutrinos determine the neutrino masses (through the traditional
seesaw mechanism), 
whereas mixings amongst the light left-handed neutrinos 
are primarily responsible for determining the effective flavor mixing angles.

It turns out that the eigenvalue phenomenology
is very different depending on whether $\delta=1$ or $\delta>1$.
For $\delta=1$, we have the mass matrix given in Eq.~(\ref{tildeM2}).
However, in this case the effective brane/bulk coupling $m$ is necessarily larger 
than the effective brane/brane components $m_{\alpha\beta}$ by a factor
$(mR)^{-1}\gg 1$.  Thus, the mixing with the bulk zero mode is dominant
and we must actually invert this ($n_{\rm f}+1$)-dimensional
mass matrix in order to examine its mass eigenvalues.
We find that in general, $n_{\rm f}-1$ eigenvalues will be of (the smaller) size
$M_\ast^2\sigma$, while the remaining two eigenvalues will be of (the larger) size $m$;
likewise, the eigenstates corresponding to $\lambda\sim M_\ast^2\sigma$ 
will have an extremely small component involving the right-handed bulk neutrino,
while the eigenstates corresponding to $\lambda\sim m$ will have
a fairly significantly component involving the right-handed bulk neutrino,
with probability $\approx \half$. 
Thus, the $n_{\rm f}-1$ lightest neutrino components 
are easily interpretable as our light flavor neutrino mass
eigenstates --- with significant flavor mixings amongst them --- 
provided we choose $n_{\rm f}$ to {\it exceed}\/
our eventual desired number of light neutrinos by one.

Given these results, we see that our scenario for $\delta=1$
actually requires an {\it additional}\/ left-handed neutrino
on the brane which participates in the flavor oscillations
but is otherwise exceedingly heavy.
In some sense, this neutrino may be regarded as an additional ``sterile'' neutrino,
since its flavor index need not correspond to any of the 
flavors exhibited by the charged fermions of the Standard Model;
moreover, by choosing an appropriate location for this neutrino,
its flavor mixings with the other left-handed neutrinos can be minimized
and/or eliminated.
However, the unique prediction of this model (as opposed to other higher-dimensional
models) is that this sterile neutrino is
necessarily {\it left-handed}\/, living on the brane.
Of course, its mass is of the same scale ($\sim m$) as that of the right-handed 
zero-mode neutrino living in the bulk.  Note that the remaining KK bulk excitations
have masses scaling as $\sim n/R$, and thus their masses are even heavier by
an additional hierarchical factor of $(mR)^{-1}\gg 1$.

For $\delta>1$,
by contrast,
the brane/bulk coupling $m$ is suppressed by 
additional factors of $(\sigma/R)\ll 1$
relative to the brane/brane components $m_{\alpha\beta}$.
Thus, 
assuming that $\sigma/R\ll (mR)^{2}$,
we see that the brane/bulk coupling will be 
extremely {\it small}\/ compared to the brane/brane components,
leading to $n_{\rm f}$ eigenvalues
of mass $\sim M_\ast^2\sigma$  and a single eigenvalue of
mass $\sim \sigma/R^2$.
Although this single mass eigenstate will be much lighter than the $n_{\rm f}$
active eigenstates, it will be overwhelmingly sterile and thus will not
interact with the Standard-Model particles.
 
We see, then, that regardless of the number of extra spacetime 
dimensions, our active neutrino mass eigenstates will have masses
which scale as $\sim M_\ast^2 \sigma$ in the current setup.
There are therefore a number of values for $M_\ast$ and $\sigma$
which can yield neutrino masses in a phenomenologically acceptable
range.  One appealing possibility, for example, is to take
$M_\ast$ (the fundamental mass scale in our theory)
of order $M_\ast\sim {\cal O}({\rm TeV})$,
while our three-branes have widths of
order $\sigma\sim {\cal O}(M^{-1}_{\rm Planck})$.
By contrast, the only solution for which 
$M_\ast$ and $\sigma^{-1}$ are of the same order of
magnitude requires that we take $M_\ast$ in the sub-eV range.
Likewise, if we demand $\sigma \sim {\cal O}({\rm TeV}^{-1})$,
then we must take $M_\ast\sim 
  {\cal O}({\rm keV})$--${\cal O}({\rm MeV})$.
However, there are   
other ways of adjusting 
the overall mass eigenvalues in this scenario
without affecting the flavor mixing angles.
For example,  
if there are two extra dimensions whose coordinates are $y$ and $z$,
we can easily imagine that the Standard-Model brane is located at $z=z_1$
while the right handed neutrino  has bulk excitations in the $y$-direction but
is localized around $z=z_2$.
If both localizations have common width $\sigma$,
we can retrace our previous steps to find that 
this location ``mismatch''
in the second extra dimension 
has the effect of inserting an extra suppression factor
\beq
   \exp\left\lbrack - {\pi\over 2}\, {(z_1-z_2)^2\over \sigma^2}\right\rbrack
\eeq
for each entry in the resulting effective mass matrix ${\cal M}$.
While such an overall rescaling of the mass matrix does not affect
the magnitudes of the flavor mixing angles, it directly suppresses 
the magnitudes of the mass eigenvalues.
Thus, in this way, we are free to adjust our mass eigenvalues as needed
without disturbing the flavor structure of our model.
Other similar scenarios are also possible.

\section{Discussion}
\label{discussion}

In this paper, we have presented a ``hybrid'' model in which the 
split-fermion scenario and the bulk-neutrino scenario are joined together.
As we have seen, this joining has succeeded in 
yielding a rich spectrum of flavor mixing angles without significant 
fine-tuning, thereby ameliorating some of the difficulties inherent
in either approach alone.  

Needless to say, we have pursued only the most ``minimal'' 
approach towards constructing this hybrid model.
Further possibilities that we have not exploited include, for example,
possible variable widths of the Gaussians~\cite{Hung}, 
the possibility of several right-handed neutrinos, 
possible orbifold twists~\cite{Harnik},
or possible additional flavor-dependent Yukawa couplings of order one.  
We have also avoided primordial flavor mixings on the brane,
and we have taken our brane neutrinos to have vanishing Majorana masses. 
While all of these assumptions can be relaxed, our main purpose in this paper
has been to illustrate the range of theoretical  
possibilities that emerge solely from the joining of the split-fermion
and bulk-neutrino models.  

Likewise, we have not performed a detailed fit of the parameters involved,
nor have we tried to embed this neutrino sector of the Standard Model
within a larger split-fermion framework that also involves the 
quarks and charged leptons.
However, we caution that in order to investigate whether such a hybrid 
model can accommodate all of 
the observed neutrino-oscillation properties, 
it is generally not sufficient merely to 
focus on reproducing the preferred mass differences and mixing angles often quoted in the
neutrino literature.
This is because such neutrino mixing parameters are usually quoted 
within the theoretical context of a simple two- or three-state neutrino mixing.
In higher-dimensional neutrino models, by contrast, one usually faces
very complex, multi-component neutrino oscillations with a variety of varying oscillation
lengths and magnitudes (see, {\it e.g.}\/,  Ref.~\cite{ds} 
for examples of the complexities involved).
Indeed, even in the small-coupling limit $mR\ll 1$, 
one typically has a four-neutrino (or multi-neutrino) oscillation,
depending on the number of bulk neutrinos which are present.
Thus, in order to perform a proper comparison with experiment,
one must go back to the original
oscillation data, plotted in terms of appropriate baselines, and perform a direct fit
between predicted and observed oscillation patterns (and suitable time-averages thereof) 
without regard for pre-existing theoretical prejudices.
This too represents a direction for further investigation.

There are also several important theoretical issues which 
we have not examined in this paper.
One pressing issue,
for example, concerns how such hybrid settings
can arise on a fundamental level. 
Although mechanisms of trapping fermions inside a fat brane have been examined 
closely, and although various 
ways to realize a split-fermion  scenario seem to be possible,
it would be interesting to have a more complete picture 
in which the required
fermion locations can be explained from some 
deeper principle and/or generated dynamically. 

Nevertheless, we have seen in this paper
that joining the split-fermion and bulk-neutrino scenarios
has produced a rich theoretical model which naturally gives rise to
large neutrino flavor mixing angles without significant fine-tuning.
Indeed, it is the coupling to the KK modes of the bulk neutrino which eliminates
the usual fine-tuning of the split-fermion scenario, and which
produces the relatively large mixing angles that we have observed.
Thus, within the context of the split-fermion scenario,
we have a natural explanation for the 
somewhat surprising fact that
quark mixing angles are generally small while lepton mixing angles are 
apparently large:  while the lepton sector contains a right-handed neutrino
which is a gauge singlet and can therefore propagate in the bulk,
all of the fields of the corresponding quark sector 
carry gauge charges and thus must live within the brane.
Thus only the lepton sector is capable of receiving the 
enhancement to the flavor mixing angles that comes indirectly 
from the couplings to the KK modes of the right-handed neutrino.
We believe that this observation is critical, and perhaps provides
one of the strongest motivations for considering such higher-dimensional
models of neutrino phenomenology.

\section*{Acknowledgments}

This work is supported in part by the National Science Foundation
under Grant PHY/0301998, by the Department of Energy under 
Grant~DE-FG02-04ER-41298,
and by a DFG grant from the German government. 
We are happy to thank Z.~Chacko, H.-S.~Goh,
and S.~Scherer for discussions.


\begin{appendix}

\section*{Appendix A:  ~Perturbative diagonalization of the mass matrix}

The neutrino mass matrix ${\cal M}$ shows a typical structure which is common for 
seesaw scenarios.
We can therefore analyze some of its features by similar means. 
Following standard treatments, 
let us write ${\cal M}$ in the block form
\beq
      {\cal M} ~=~  \pmatrix{ 
         M_L &  M_D    \cr
         M_D^T   &  M_R   \cr }
\eeq
where the $M_L$ block corresponds to the brane neutrinos as well as the
zero mode of the bulk neutrino, where the $M_R$ block corresponds to the excited KK modes
of the bulk neutrino, and where $M_D$ corresponds to the couplings between these two
groups of states. 
Assuming that these couplings are small ({\it i.e.}\/, assuming
$mR\ll 1$, so that we are in the ``perturbative'' regime), we can 
then approximately block-diagonalize this matrix
by finding a matrix $U$ such that
\beq
         U^T {\cal M} U ~=~ \pmatrix{
              \widetilde M_L & 0 \cr
              0  & \widetilde M_R \cr}~.
\eeq 
In general, the solution for $U$ is given by
\beq
          U~=~ \pmatrix{  1 & \kappa \cr 
                       - \kappa^T & 1 \cr }~~~~~~
       {\rm with}~~ 
         \kappa ~\approx~ M_D M_R^{-1}~,
\eeq
whereupon we find
\beq
       \widetilde M_L ~\approx~ M_L - M_D M_R^{-1} M_D^T~,~~~~~ 
       \widetilde M_R ~\approx~ M_R ~.
\eeq
Since our original mass matrix is real and symmetric,  
the eigenvalues are necessarily real.

The next-order corrections have the form 
$M_D M_R^{-1} M_D^T M_D M_R^{-1}$,
and an inspection shows that
these are suppressed by a factor of $ m R$ relative 
to those considered above. 
Likewise, the mixing between the excited KK modes 
and the brane/zero-mode subsector 
is suppressed by a factor of $m^\alpha_{n,\pm} R \ll 1$.

Note that in the above, we have chosen to group our blocks in such a way
that the zero mode of the bulk neutrino is joined with the brane neutrinos
rather than with the bulk-neutrino excited KK states.
This is because the KK zero mode is the only mode from the KK tower which
is not heavy, and which therefore fails to decouple.  By arranging our
blocks in this manner, we are therefore ``integrating out'' only the excited
KK modes but retaining the zero mode in our low-energy reduced mass
matrix $\widetilde M_L$.

Given the parameters in our specific model,
a straightforward calculation gives
\beq
    \kappa ~=~ \pmatrix{
             m_-^{\alpha T} & m_+^{\alpha T}\cr
                      0     &   0 \cr} \, {R\over n}~,
\eeq
whereupon
we find
\beqn
     \kappa M_D^T &=& \pmatrix{
             m_-^{\alpha T} & m_+^{\alpha T}\cr
                      0     &   0 \cr} \, {R\over n}\,
                 \pmatrix{
               m_+^\beta & 0 \cr
               m_-^\beta & 0 \cr }\nonumber\\
          &=&  \pmatrix{
             -m_{\alpha \beta} & 0 \cr
            0        & 0 \cr}
\eeqn
with
\beq
           m_{\alpha \beta}  ~=~ 
           -\sum_{n=1}^{\infty} \left( m^{\alpha T}_{n,+} \, \frac{R}{n}\, m^\beta_{n,-} ~+~     
           m^{\alpha T}_{n,-}  \, \frac{R}{n}\, m^\beta_{n,+} \right) ~.
\label{malphabeta}         
\eeq
We thus find
\beq
       \widetilde{M}_L~\approx~ \pmatrix{ 
             m_{\alpha \beta}    & m  \cr
                     m^{T}&0  \cr}~,
\eeq
whereupon inserting the definitions
in Eq.~(\ref{couplings3}) yields the result given in
Eq.~(\ref{interm}).

\section*{Appendix B:  ~Mass Scales for $\delta>1$}

We now consider the mass scales involved in our mass matrix $\widetilde {\cal M}$.
We are particularly interested in the mass scales associated with the expression $m_{\alpha\beta}$
in Eq.~(\ref{malphabeta}).
Thus, there are three components of this expression that we need to generalize
to $\delta$ dimensions:
we need to generalize the couplings $m^\alpha_{n,\pm}$;
we need to generalize expressions such as $1/n$;
and we need to generalize the KK sum.
We shall therefore begin by addressing each of these in turn. 

For excitations of the bulk field in more than one extra spacetime dimension, the Kaluza-Klein
expansion in Eq.~(\ref{KKdecomp}) will now involve products of sines and cosines.
This in turn means that the couplings $m^\alpha_{n,\pm}$
in Eq.~(\ref{couplings3})
will also be products of sines and cosines whose
arguments display the positions $y_i$ of our fields.

Likewise, the form of the higher-dimensional Clifford algebra implies that
a Dirac spinor in $4+\delta$ spacetime dimensions will have
$2^{2+\delta/2}$ components, and the inverse of the wave vector $n$ will be a
linear combination of its components.
However,
since our goal is merely to estimate relevant mass scales,
we shall ignore these spinorial details
and retain a notation whereby we treat quantities such as $1/n$ as a single number.

Finally, we observe that
the KK sum in Eq.~(\ref{malphabeta}) will now extend over all $\delta$-dimensional
vectors $n=(n_1,n_2,...,n_\delta)$ with $n_i\in \IZ^+$ except
the zero vector.

Let us now put these ingredients together.
First, we notice that in the higher-dimensional case with $\delta>1$, 
the couplings from Eq.~(\ref{KKdecomp}) will enter into Eq.~(\ref{fdef}) only for 
mixtures of even and odd fields. This means
that the factors that must be summed as part of the $n$-summation 
will consist of at least one sine function and one cosine
function.
Since the contributions to this sum are naturally damped 
beyond $n \sim R/\sigma$, and
since we assume the locations $y_\alpha$ to 
be of order $\sigma$, we can approximate each occurrence of the sine function
by the linear term in its Taylor expansion
(and set each cosine factor to $1$).
Let $s$ denote the number of sine functions
in the product, where $1<s<\delta$, and likewise let $i=1,...,s$ denote
those directions for which these sine functions appear.  
Each of these sine functions will then make a linear contribution 
to $m_{\alpha\beta}$
of the form $y_i n/R$,
whereupon we find that Eqs.~(\ref{8ev}) and (\ref{fdef})
will generalize to yield an expression of the form 
\beq
    m_{\alpha \beta}  ~\sim~  m^2 R\,
    \sum_n \frac{1}{n} \, \exp \left( - \frac{n^2 \sigma^2}{\pi R^2} \right)\, 
          \prod_{i=1}^s \left( y_i \frac{n}{R} \right)~.
\eeq
Note that this $n$-summation continues to represent a sum over $\delta$-dimensional vectors $n$. 
Defining $k\equiv (\sigma/R) n$, we see that this sum 
can be approximated by an  integral
\beq
        m_{\alpha \beta}  ~\sim~   {m^2 R^\delta \over \sigma^{\delta-1}}\, 
             \prod_i \left( y_i/\sigma\right)\, 
        \int d k~ k^{\delta +s-2}  \exp \left( - k^2/\pi^2\right)~.
\label{step}
\eeq
However, the integral is now purely dimensionless and $\sim {\cal O}(1)$;  likewise, for $y_i\sim \sigma$, 
we see that each of the product factors are also $\sim {\cal O}(1)$.
Thus the overall mass scale is set by the prefactor in Eq.~(\ref{step}).  However,
given the result in Eq.~(\ref{replace}),
we see that we can write this prefactor in the form $M_\ast^2 \sigma$, obtaining the same overall
mass scale as we found for $\delta=1$.
Thus, we conclude that the overall mass scale for $m_{\alpha\beta}$ is 
independent of both $R$ and $\delta$.
Indeed, it is easy to verify that this result remains true even when all of the
proper numerical and spinorial factors are included in the analysis.

\end{appendix}


\end{document}